\documentclass[%
reprint,
showpacs,preprintnumbers,
aps,
prc,
]{revtex4-1}
\usepackage{graphicx}
\usepackage{dcolumn}

\begin{document}
\newcommand{\Cu}[1]{$^{#1}\mathrm{Cu}$}

\title{Simulation of $^{67}$Cu photo-nuclear production in nanoparticles}

\author{D.V. Fedorchenko}
\email[Corresponding author: ]{fdima@kipt.kharkov.ua}

\author{M.A. Khazhmuradov}
    
\author{Y.V. Rudychev}
\affiliation{National Science Center ``Kharkov Institute of Physics and
        Technology'', Kharkiv, Ukraine}    
    
\begin{abstract}
  The process of
  $^{67}\mathrm{Cu}$ nuclide photoproduction in the zinc dioxide nanoparticles
  immersed in the water media was simulated.  We calculated the escape fractions 
  of $^{67}\mathrm{Cu}$ nuclei and corresponding ranges in water for nanoparticle 
  sizes from 40~nm to 80~nm and incident photons energies from 12~MeV to 30~MeV. 
  Usage of capturing nanoparticles for accumulation of the escaped $^{67}\mathrm{Cu}$ 
  nuclei is also discussed.
\end{abstract}
    
    
\maketitle

\section{Introduction}
During the last decades \Cu{67} nuclide is a subject of numerous investigations
as a perspective candidate for treatment and diagnostics of various types of cancer \cite{Schubiger2002,   Smith2012}. The permanent interest to this isotope is motivated by its' attractive radiological characteristics. \Cu{67} nuclide has low toxicity and exhibits no tendency to accumulate in bones or organs. The half-life period of 2.58 days provides the necessary therapeutic irradiation doze. \Cu{67} nuclide emits $\beta^{-}$ particles with weighted average energy of 141~keV.  The corresponding $\beta^{-}$ particle range of 0.2~mm in tissue makes \Cu{67} suitable for the treatment of small tumors \cite{Schubiger2002}. The \Cu{67} nuclide also has $\gamma$-emissions at 91.266~keV (6\% abundance), 93.311~keV (35\% abundance) and 184.577~keV (45\% abundance) that can be used for diagnostic purposes.

The main sources of the \Cu{67} nuclide are nuclear reactions
on zinc through the (p,2p), (n,p) and ($ \gamma $,n) production channels.  The production method using $^{68}\mathrm{Zn}(\mathrm{p},2\mathrm{p}){}^{67}\mathrm{Cu}$ reaction on the high-energy proton accelerators provides a rather high \Cu{67} yield but is accompanied by the production of considerable amounts of unwanted isotopes \cite{Smith2012}. As a result a complex separation scheme is necessary to obtain the required radiological purity \cite{Dasgupta1991, Medvedev2012}.  Another known production method uses neutron-induced reaction $^{67}\mathrm{Zn}(\mathrm{n},\mathrm{p}){}^{67}\mathrm{Cu}$ in a
nuclear reactors \cite{Mirzadeh1986,Mausner1998}. The complications associated with this method include activation of the target containment vessel and accompanied production of \Cu{64} isotope through the 
$^{64}\mathrm{Zn}(\mathrm{n},\mathrm{p}){}^{64}\mathrm{Cu}$ reaction
\cite{Mirzadeh1986}. 

Photonuclear method of \Cu{67}
production uses the 
$^{68}\mathrm{Zn}(\gamma,\mathrm{p}){}^{67}\mathrm{Cu}$ reaction
induced by the bremsstrahlung radiation on the electron linacs
\cite{Malinin1970,Aizatsky2010,Starovoitova2014,Howard2015}. This method also suffers from the production of unwanted isotopes, but the effect could be reduced by using the enriched target \cite{Kondo1978}. Photonuclear method is considered the  most promising  as it provides the \Cu{67} yield comparable to the yield of neutron-induced reaction and higher than that for the proton-induced reaction. At the same time less amount of waste is produced and the eventual product has higher radiological purity. 

In this paper we consider the photonuclear method for \Cu{67} production. The typical setup for \Cu{67} production consists of electron accelerator, bremsstrahlung converter and production target. The conventional target is a solid body made from metal zinc or zinc oxide~\cite{Aizatsky2010,Howard2015}. Practical usage of such target encounters two main problems: high heat deposition rate and low specific activity after the irradiation cycle.  Thus the \Cu{67} production setup requires efficient cooling system to prevent target damage and  also must be followed by a complicated extraction process to achieve reasonable yield of \Cu{67} nuclide.

The possible way to reduce the heat loads and increase the specific activity is to use the liquid production target containing the suspension of zinc oxide nanoparticles~\cite{Dikiy-nano}.  Production of \Cu{67} in such target has several peculiarities compared to the solid target. The most essential is kinematic recoil effect when the daughter nuclei from the photonuclear reaction escape the nanoparticle. This effect is negligible for the large solid target, but for the nanoparticle with the size of tens of nanometers the escape fraction of the recoil nuclei is rather high~\cite{Starovoitova2014}. Those escaped \Cu{67} nuclei are accumulated in the ambient liquid and could be subsequently extracted, while the nanoparticles could be filtered out and reused. Such approach provides efficient recycling of the production target and the decreases the amount of the resulting radioactive wastes.

In this paper we calculated the escape fractions of \Cu{67} nuclei from the zinc oxide nanoparticles the corresponding ranges in water. The spectrum of the daughter \Cu{67} nuclei was calculated using Talys-1.8 program code~\cite{Talys1} and for the simulation of \Cu{67} nuclei transport we used GEANT4.10.2p01 toolkit~\cite{Geant1,Geant2} with high precision model for ion transport.

\section{Methods}

\subsection{\label{sec:recoil} Calculations of the recoil spectra}

\begin{figure}
    \includegraphics{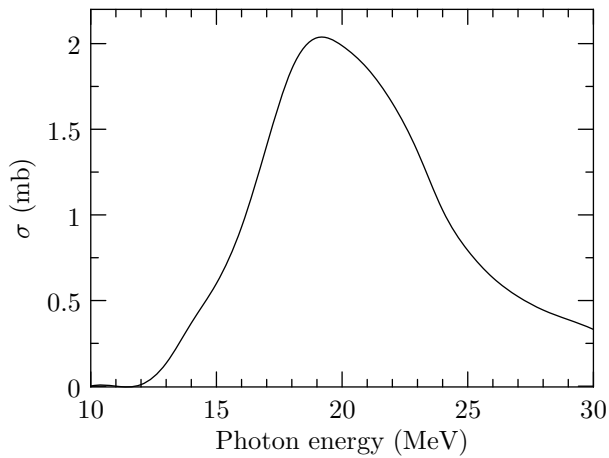}
    \caption{\label{fig:talys-cs}Cross section of the $^{68}\mathrm{Zn}(\gamma,\mathrm{p}){}^{67}\mathrm{Cu}$ reaction calculated using Talys-1.8~\cite{Talys1}}
\end{figure}

The most essential aspect in the calculation of the escape fraction and range of \Cu{67} nuclei in nanoparticle is simulation of the $^{68}\mathrm{Zn}(\gamma,\mathrm{p}){}^{67}\mathrm{Cu}$ reaction. This reaction has a broad maximum centered at 19~MeV, that corresponds to the Giant Dipole Resonance (GDR) effect (see~\ref{fig:talys-cs}). One can see the total cross section of this process is rather low -- about 2~mb at GDR maximum. Estimation of the probability of the ($ \gamma $,p) reaction for 40~nm zinc oxide nanoparticle with this cross section gives the value of $1.11\cdot 10^{-12}$. Such low interaction probability makes the direct Monte Carlo simulation of the photons interaction with the  individual nanoparticle highly inefficient. 

\begin{figure}
    \includegraphics{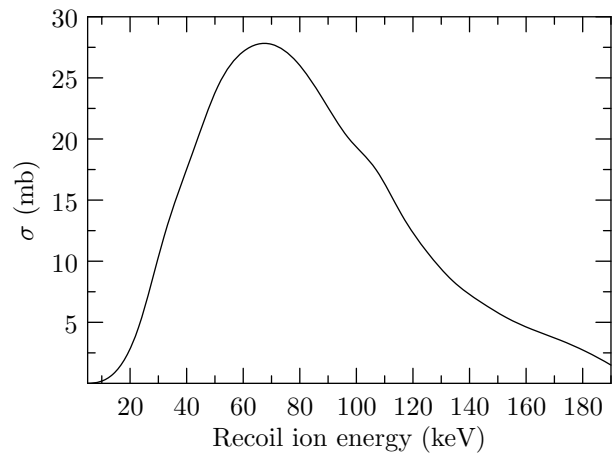}
    \caption{\label{fig:talys-rec}Recoil spectrum of the \Cu{67} ions for 20~MeV incident photons calculated using Talys-1.8~\cite{Talys1}}
\end{figure}

For the calculation of the escape fraction and range of \Cu{67} nuclei one needs only the kinematic characteristics of the daughter nuclei after the photonuclear reaction. Thus, to avoid the time-consuming Monte Carlo simulations we used the Talys code to obtain the energy spectra of the \Cu{67} recoil nuclei for different energies of the incident photons. The example spectrum for 20~MeV incident photons is shown in the Fig.\ref{fig:talys-rec}. The shape of the calculated spectrum has essential differences from the evaporation model, namely the amount of high-energy recoil nuclei (and corresponding high-energy protons) is significantly higher. This agrees with the results of the experimental studies~\cite{Osokina1967}; the increased fraction of high-energy recoils could be attributed to the direct photo-nuclear reactions. 

\begin{figure}
    \includegraphics{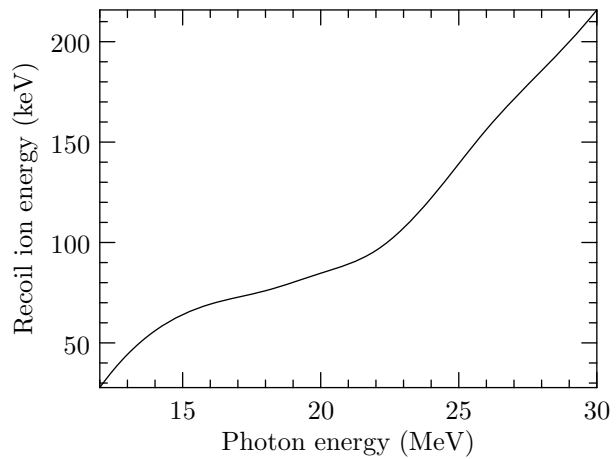}
    \caption{\label{fig:talys-en}Average energies of the \Cu{67} ions for 20~MeV incident photons calculated using Talys-1.8~\cite{Talys1}}
\end{figure}

Figure~\ref{fig:talys-en} represents the average energies of the recoil \Cu{67} nuclei for the incident photon energies between 12~MeV and 30~MeV. For the photons with energies around GDR peak that make the most essential contribution to the \Cu{67} production the average kinetic energies of the recoil nuclei are of 60-130~keV. Such high-energy recoil nuclei have a rather high probability to escape from the small sized nanoparticle.

\begin{figure}
    \includegraphics{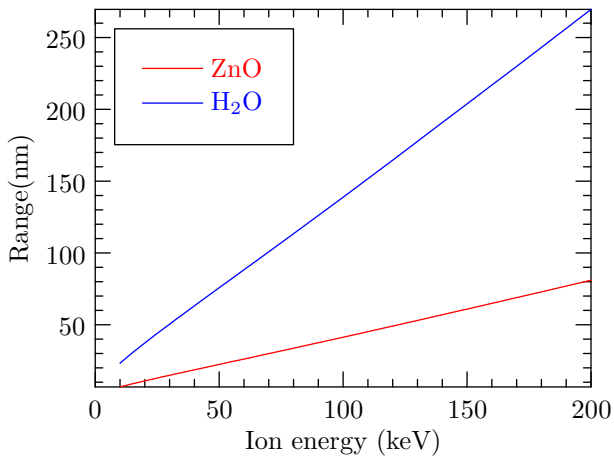}
    \caption{\label{fig:srim-ranges}Ranges of \Cu{67} ions calculated using SRIM~\cite{SRIM}}
\end{figure}

To assess the escape probability we calculated the range of \Cu{67} ions in zinc oxide and water using the SRIM codes\cite{SRIM}. The corresponding dependencies presented in the Figure~\ref{fig:srim-ranges} show that \Cu{67} ions range in zinc oxide is the order of magnitude of the nanoparticle size. The crude estimation gives the escape probability of approximately 50-60\% for photon energies around 20~MeV.

SRIM(TRIM) codes are also capable to perform accurate Monte Carlo simulation of ion transport~\cite{SRIM}. However,  this package has very limited capabilities on the geometry being simulated: only plain geometries such as layers with constant thickness are allowed. Thus, for simulation of the \Cu{67} nuclei transport and calculation of  escape fraction we used the GEANT4 package with the recoil spectra calculated using Talys~1.8 codes.

\subsection{Simulation of ion transport in the nanoparticle and in ambient media}

The recoil \Cu{67} nucleus experiences elastic collisions with the atoms of the nanoparticle and  with  atoms of the ambient media. Collisions with the atoms of zinc in the nanoparticle volume effectively decelerate \Cu{67} nucleus because of almost equal atomic masses of zinc and copper nuclei. Thus, accurate simulation of the elastic scattering process is of crucial importance for calculation of escape fraction and range of \Cu{67} nuclei.

Within the standard approach GEANT4 treats the elastic ion scattering as continuous multiple scattering process (MSC). The MSC simulation uses several approximate models~\cite{Ivanchenko2010} based on statistical description of the scattering process. While that models provide sufficient accuracy for thick target where the number of collisions is hight and statistical approach is valid, small-sized 
nanoparticles require exact simulation of individual scattering events to obtain correct results.

The accurate GEANT4 model for elastic ion scattering was developed by Mendenhall and Weller~\cite{Mendenhall2005420}. Their approach is based on classical scattering integral for the two-particle scattering with screened Coulomb potential. Such calculation scheme provides perfect agreement with SRIM simulations and experimental data on ion transport in thin foils. To include this model into GEANT4 code we used \textsf{G4ScreenedNuclearRecoil} class from the examples included in GEANT4 code bundle.

For the actual simulations of the \Cu{67} recoil nuclei transport we used simple geometry setup: spherical zinc oxide nanoparticle ($ \rho = 5.61\ \mbox{g}/\mbox{cm}^3$) immersed into water media. To simulate $^{68}\mathrm{Zn}(\gamma,\mathrm{p}){}^{67}\mathrm{Cu}$ reaction we generated \Cu{67} ions at random points inside the nanoparticle volume, and the energy of the generated ions was sampled from the recoil energy spectra calculated using Talys code (see section~\ref{sec:recoil}). Experimental data on photoproton reaction on zinc shows weak anisotropy in the angular distribution of the reaction products~\cite{Osokina1967}, however during our simulations we neglected it and momentum direction of the generated \Cu{67} ions was sampled isotropically. During simulation  $ 10^6$ initial \Cu{67} nuclei were generated to achieve the reasonable statistics of the calculated values.

\section{Results and Discussions}

\begin{figure}
    \includegraphics{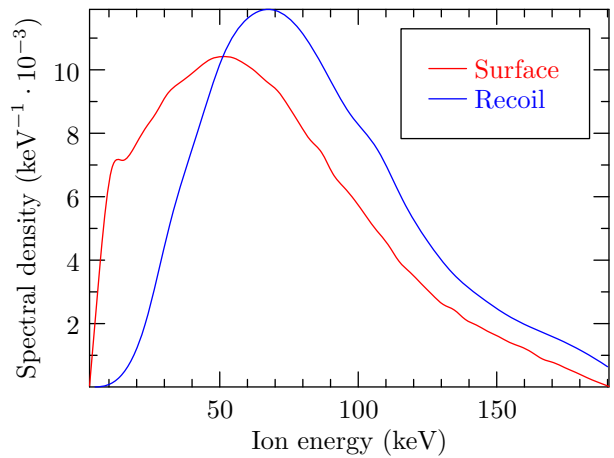}
    \caption{\label{fig:recoil-spectra}Spectrum of outgoing \Cu{67} ions at nanoparticle surface and initial recoil spectrum for 40~nm nanoparticle and 20~MeV incident photons}
\end{figure}

\begin{figure}
    \includegraphics{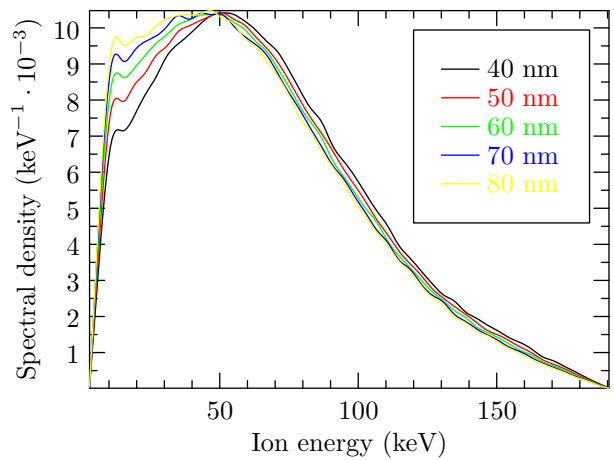}
    \caption{\label{fig:recoils-all}Spectrum of outgoing \Cu{67} ions at nanoparticle surface for  20~MeV incident photons}
\end{figure}

The simulations showed essential influence of the elastic collisions inside the nanoparticle on the \Cu{67} nuclei transport. Figure~\ref{fig:recoil-spectra} shows the spectrum of the outgoing \Cu{67} ions at the nanoparticle surface compared to the initial recoil spectrum. The elastic collisions widen the energy peak and shift it to the lower energies. This effect becomes more pronounced for big nanoparticles, as the number of elastic collisions increases (see Fig.~\ref{fig:recoils-all}).  

\begin{figure}
    \includegraphics{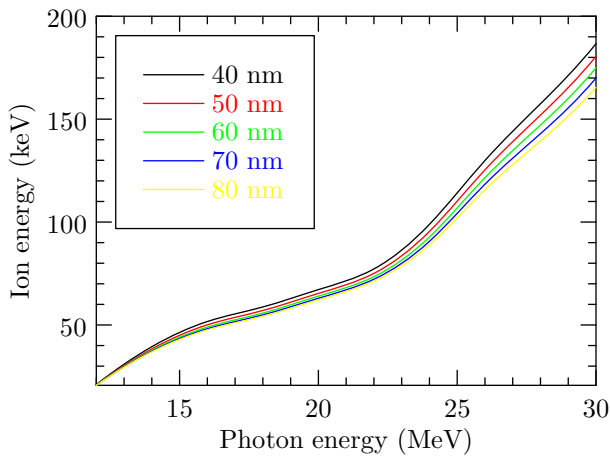}
    \caption{\label{fig:energies}Average energies of the escaping \Cu{67} ions at the nanoparticle surface}
\end{figure}

In the section~\ref{sec:recoil} we have mentioned that there are two mechanisms of photoproton reaction on zinc: photon capturing followed by the compound nucleus stage and direct nuclear photo-effect. The first produces outgoing protons with evaporation spectrum, while direct reactions produce high-energy protons. The contribution from the direct reactions increases with the energy of the incident photons, and the average energies of outgoing protons and recoil nuclei also increase. The calculated average energies of the outgoing \Cu{67} nuclei at the nanoparticle surface are presented in the Figure~\ref{fig:energies}. One can see the average energy of escaping ions exceeds 200~keV for 30~MeV photons. Such high energy ions cause substantial ionization of the hydrogen and oxygen atoms of the ambient water. The  active hydrogen and oxygen could enter into reaction with \Cu{67} ions forming  chemically bound compounds. 
This effect could cause complications during the extraction process and lower the eventual yield of \Cu{67} nuclide. Obviously, ionization effects could be mitigated by limiting the energy of the incident photons, but at a cost of a decreased yield. Another possible solution is usage of admixture of chemically inert nanoparticles that capture outgoing \Cu{67} ions.

\begin{figure}
    \includegraphics{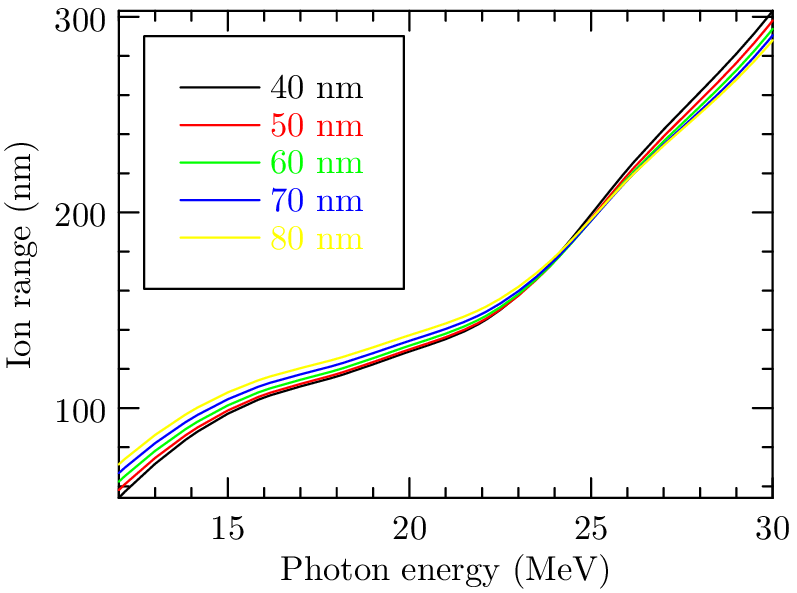}
    \caption{\label{fig:ranges}Ranges of the \Cu{67} ions in water}
\end{figure}

The required concentration of these capturing nanoparticles could be estimated from the \Cu{67} ion range in water. Calculated ranges for incident photon energies up to 30~MeV are shown in the Figure~\ref{fig:ranges}. Taking into account that maximum production of \Cu{67} corresponds to the incident photon energy around 19~MeV (GDR peak, see Figure~1), we assumed the value of $R_{\mbox{Cu}}\approx 110\ \mbox{nm}$ as a characteristic \Cu{67} ion range. To capture the escaping \Cu{67} ions at least one capturing nanoparticle must be present inside the sphere with radius $ R_{\mbox{Cu}} $ around the zinc oxide nanoparticle. This gives the crude estimation of the lower boundary for concentrations of capturing nanoparticles $ n_{capture} $ and zinc oxide nanoparticles $ n_{nano} $ of $ n_{nano} \ge n_{capture} \approx (2R_{\mbox{Cu}})^{-3} = 9.4\cdot 10^{19}\ \mbox{m}^{-3} $. Obviously, the concentration of the producing zinc oxide nanoparticles must be essentially higher to ensure the efficient capturing of the escaping \Cu{67} ions.

\begin{figure}
    \includegraphics{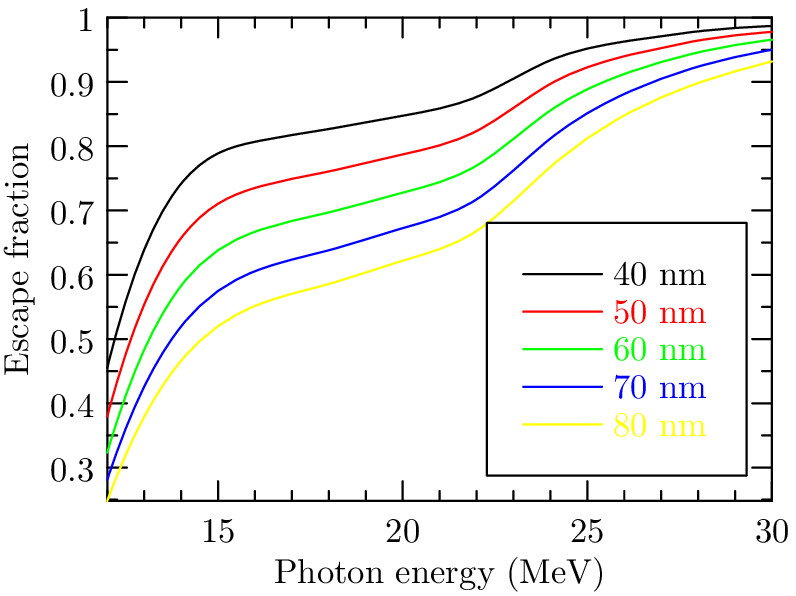}
    \caption{\label{fig:yields}Ranges of the \Cu{67} ions in water}
\end{figure}

Another important parameter for the assessment of \Cu{67} production in the nanoparticles is escape fraction of \Cu{67} nuclei. We defined the escape fraction as a ratio of the number of \Cu{67} ions passing through the nanoparticle surface outside and the number of generated initial ions. The results of the simulations for the nanoparticle sizes from 40~nm to 80~nm are shown in the Figure~\ref{fig:yields}. 

This calculated values agree with the crude estimations based on SRIM calculations of \Cu{67} ranges (see \ref{sec:recoil}). From the calculated dependencies it follows that for GDR peak at 20~MeV escape fractions vary from 0.62 for 80~nm nanoparticles to 0.85 for 40~nm nanoparticles. This difference persists for photon energies corresponding to the GDR region. For higher energies due to the saturation effect escape fraction goes close to the maximum value of 1. The further increasing of the incident photon energy will not provide in higher \Cu{67} yield, but will have detrimental effect due to high ionization of the ambient water as was mentioned above. 

\section{Conclusions}

We performed calculations of the \Cu{67} nuclei escape fraction from the zinc oxide nanoparticles and corresponding ranges of the escaped nuclei in the ambient water. The two stage calculation scheme was implemented: the recoil spectra of the \Cu{67} nuclei were calculated using Talys-1.8 code, and then these spectra were used for Monte Carlo simulation of \Cu{67} ions transport using GEANT4 code. This approach allowed to obtain statistically valid results despite of the low cross section of $^{68}\mathrm{Zn}(\gamma,\mathrm{p}){}^{67}\mathrm{Cu}$ reaction.

Our calculations showed that escape fraction of the \Cu{67} nuclei is rather high for incident photons energies in the GDR region of $^{68}\mathrm{Zn}(\gamma,\mathrm{p}){}^{67}\mathrm{Cu}$ reaction. For 40~nm zinc oxide nanoparticles it increases from 0.80 to 0.94 for photon energies from 15~MeV to 25~MeV. For the higher photon energies saturation occurs, while the process cross section considerably decreases. Thus the eventual yield of \Cu{67} for photon energies above 30~MeV is lower than that for the photons with energies in the GDR region.

We have also considered usage of the capturing nanoparticles to collect the escaped \Cu{67} nuclei. Using the calculated range of the \Cu{67} ions in water we obtained the estimation of lower boundary value for concentration of such capturing nanoparticles of about $ 9.4\cdot 10^{19}\ \mbox{m}^{-3} $.

\end{document}